\documentclass[12pt]{article}
\usepackage[cp1251]{inputenc}
\usepackage[russian]{babel}
\usepackage{graphicx}
\usepackage{amsmath}
\usepackage{amssymb}
\usepackage{latexsym}

\begin{document}
\title{Rayleigh surface wave interaction with 2D exciton Bose-Einstein condensate}
\author{M.\,V.\,Boev, V.\,M.\,Kovalev\\Institute of Semiconductor Physics,\\ Siberian Branch of the Russian Academy of Sciences,\\ Novosibirsk, 630090, Russia\\
Novosibirsk State Technical University, Novosibirsk, 630095, Russia}

%\address{Институт физики полупроводников им.~А.В. Ржанова СО РАН, 630090 Новосибирск, Россия}
%\address{$^{\star}$Новосибирский Государственный Университет, 630090 Новосибирск, Россия}
%\address{$^{\circ}$Новосибирский Государственный Технический Университет, 630095 Новосибирск, Россия}

\date{19 September 2014}
\maketitle

\begin{abstract}
We describe the interaction of the Rayleigh surface acoustic wave (SAW) traveling on the semiconductor substrate and interacting with excitonic gas in a double quantum well located on the substrate surface.  We study the SAW attenuation and its velocity renormalization due to coupling with excitons. Both the deformation potential and piezoelectric mechanisms of the SAW-exciton interaction are considered. We focus our attention on the frequency and excitonic density dependencies of the SAW absorption coefficient and velocity renormalization at temperatures both above and well below the critical temperature of Bose-Einstein condensation of excitonic gas. We demonstrate that the SAW attenuation and velocity renormalization are strongly different below and above the critical temperature.
\end{abstract}
\section{Introduction} A gas of a bound electron-hole pairs, excitons, being the Bose-like particles, can exhibit the Bose-Einstein condensation (BEC) at extremely low temperatures. This phenomenon was theoretically predicted long time ago \cite{Moskalenko},\cite{Blatt},\cite{Keldysh_Kopaev},\cite{Keldysh_Kozlov},\cite{Keldysh_Guseinov} and was intensively studied in the $Cu_2O$ system (see recent review article \cite{Snoke}). Recently, the BEC of excitons in low dimensional systems has been confirmed in various experiments \cite{Snoke2},\cite{Butov}, \cite{Kasprzak}.

The experimental evidence of the existence of exciton BEC is mainly based on optical arguments. The general idea is the narrowing of the luminescence line when the exciton gas is cooled down to below the critical temperature.

The main aim of the present work is to theoretically demonstrate that the SAW experimental technique widely used in earlier studies of two-dimensional electron gas \cite{ChaplikKrasheninnikov}  may provide with an \emph{alternative} method to study the exciton BEC. We show that the SAW velocity renormalization $\Delta c/c$ and SAW attenuation coefficient behave differently above and below the critical BEC temperature and this may be used as an experimental confirmation of exciton BEC.

We consider the double quantum well (DQW) structure depicted in Fig. 1. The electron and hole are located in different QWs interacting via Coulomb potential forming the exciton with the dipole moment $\textbf{p}$ directed along the normal of the DQW plane. We will assume the simple exciton model. Exciton will be considered as a rigid dipole particle with a dipole moment along direction $z$ only, $\textbf{p}=(0,0,-ed)$. Here $e$ is an electron charge and $d$ is a distance between QWs. Such a model ignores the internal motion of the particles and the motion in the $z$-direction. This model is good enough to describe the system under study, while the internal degrees of freedom are not excited. We assume that time-dependent acoustic and electric SAW fields cannot excite them. Nevertheless, the dipoles, as a whole, are free to move in the $(x,y)$ plane. SAW may interact with excitonic gas via either deformation potential or piezoelectric mechanisms. Acoustic and electric SAW fields are assumed to be the perturbations disturbing excitonic gas from the equilibrium. The response of excitonic gas to the SAW perturbation depends on whether it is in the BEC state or not resulting in different behavior of the SAW velocity renormalization and attenuation coefficient. We will consider the Rayleigh wave and start with the deformation potential mechanism.

\section{SAW-exciton interaction via deformation potential} We assume that the substrate is an isotropic elastic medium. The Rayleigh wave traveling along the surface is characterized by transverse $c_t$ and longitudinal $c_l$ sound velocities. Moreover, a typical SAW wavelength is much lager than the distance $d$ between QWs. In this case, the influence of excitonic gas on the SAW propagation can be described by changing the boundary conditions for stress tensor $\sigma_{ij}$ at surface $z=0$. Substrate displacement vector $\textbf{u}$ satisfies the equation
\begin{equation}
\label{eq.1}
\ddot{\textbf{u}}=c_t^2\Delta\textbf{u}+(c_l^2-c_t^2)\textrm{grad}\,\textrm{div}\,\textbf{u}
\end{equation}
and, in case of Rayleigh wave, it has $z$ and $x$ components $u_x(x,z)=u_x(z)e^{ikx-i\omega t}, u_y=0, u_z(x,z)=u_z(z)e^{ikx-i\omega t}$ \cite{Landau_Lifshits}, were
\begin{eqnarray} \label{eq.2}
u_z(z)=-i\kappa_lBe^{\kappa_lz}-ikAe^{\kappa_tz},\,\,\,
u_x(z)=kBe^{\kappa_lz}+\kappa_tAe^{\kappa_tz},\\\nonumber
\kappa_l=\sqrt{k^2-\omega^2/c_l^2},\,\,\,
\kappa_t=\sqrt{k^2-\omega^2/c_t^2}.
\end{eqnarray}
Arbitrary amplitudes $A,B$ are found from boundary conditions $\sigma_{ij}\tau_j=f_i$, where $f_i$ -- is a surface force (force per unit area) acting from the excitons upon the substrate surface and $\tau_j$ -- is a unit vector normal to surface $z=0$. Surface force $\textbf{f}$ arises due to the exciton density deviation from the equilibrium
\begin{eqnarray} \label{eq.3}
\textbf{f}=\lambda\,\textrm{grad}\,n
\end{eqnarray}
Here $\lambda=\lambda_e+\lambda_h$ -- is a sum of electron and hole deformation constants, $n=n_{k\omega}e^{ikx-i\omega t}$ -- exciton density fluctuation.

Thus, the boundary conditions at surface $z=0$ yield
\begin{eqnarray} \label{eq.4}
\rho c_t^2\left(\frac{\partial u_x}{\partial z}+\frac{\partial u_z}{\partial x}\right)=\lambda\frac{\partial n}{\partial x}\\\nonumber
c_l^2\frac{\partial u_z}{\partial z}+(c_l^2-2c_t^2)\frac{\partial u_x}{\partial x}=0,
\end{eqnarray}

Exciton density fluctuation amplitude $n_{k\omega}$ can be found using the standard linear response theory $n_{k\omega}=S_{k\omega}W_{k\omega}$, where $W_{k\omega}=\lambda(\textrm{div}\,\textbf{u})_{z=0}=\lambda(iku_x+\partial_zu_z)|_{z=0}$ is a potential energy of exciton in the SAW deformation field. The structure of response function $S_{k\omega}$ depends on the exciton gas state. Substituting $n_{k\omega}$ in the boundary conditions (\ref{eq.4}) and taking into account the equations (\ref{eq.2}), we get the dispersion equation
\begin{eqnarray} \label{eq.5}
f_{k\omega}=\frac{2(\lambda k\omega)^2\kappa_t}{\rho(c_lc_t)^2}S_{k\omega}\\\nonumber
f(k,\omega)=(\kappa_t^2+k^2)^2-4\kappa_l\kappa_tk^2.
\end{eqnarray}
In the absence of SAW-exciton interaction $\lambda=0$, SAW dispersion $\omega(k)$ is given by $f(k,\omega)=0$ and is linear in $k$: $\,\omega=c_t\xi_0 k$.  $\,\xi_0$ is a solution of equation $f(k,\omega=c_t\xi_0k)=k^4f(\xi_0)=0$, where \cite{Landau_Lifshits}
\begin{eqnarray} \label{eq.6}
f(\xi_0)=(2-\xi_0^2)^2-4\sqrt{1-\xi_0^2}\sqrt{1-\frac{c_t^2}{c_l^2}\xi_0^2}
\end{eqnarray}
Because of the interaction, $\xi_0$ has a $\delta\xi$ correction due to the presence of the r.h.s. in equation (\ref{eq.5}). $\delta\xi$ is a complex value, the real part of which describes the SAW velocity renormalization, and the imaginary part gives SAW attenuation coefficient $\Gamma$
\begin{eqnarray} \label{eq.7}
\frac{\Delta c}{c}=Re\,\left(\frac{\delta\xi}{\xi_0}\right),\,\,\,\Gamma=-2kIm\,\left(\frac{\delta\xi}{\xi_0}\right),
\end{eqnarray}
here $c=c_t\xi_0$ and $\omega_0=ck$.

To find $\delta\xi$, we substitute $\omega=c_tk(\xi_0+\delta\xi)$ in equation (\ref{eq.5}), expanding $f(\xi_0+\delta\xi)\approx f(\xi_0)+f'(\xi_0)\delta\xi$, and solve it by successive approximation. The result reads
\begin{eqnarray} \label{eq.8}
\delta\xi=\frac{2(\lambda \xi_0)^2\sqrt{1-\xi_0^2}}{f'(\xi_0)\rho c_l^2}kS_{k,\omega=c_t\xi_0k}
\end{eqnarray}
Thus, we can see from (\ref{eq.8}) that the imaginary and real parts of $\delta\xi$ are determined by response function $S_{k,\omega}$. To find it, we consider cases $T>T_c$ and $T<T_c$ below separately. $T_c$ is the exciton gas condensation temperature.

\section{$T>T_c$} At a high temperature and low density, the excitons can be considered as a weakly interacting gas. The interaction potential is nothing but the repealing dipole-dipole interaction. The Fourier transform of interaction potential is
\begin{eqnarray} \label{eq.9}
g(k)=\frac{4\pi e^2}{(\epsilon+1)k}\left(1-2e^{-kd}\right)+\frac{2\pi e^2}{\epsilon k}\left(1+\frac{\epsilon-1}{\epsilon+1}e^{-2kd}\right),
\end{eqnarray}
where $\epsilon$ is a QWs dielectric constant. Using the mean field approach, we find the response function
\begin{eqnarray} \label{eq.10}
S_{k\omega}=\frac{\Pi_{k\omega}}{1-g_k\Pi_{k\omega}},\\\nonumber
\Pi_{k\omega}=\sum_{\bf{p}}\frac{f^B_{\bf{p+k}}-f^B_{\bf{p}}}{\omega+E_{\bf{p+k}}-E_{\bf{p}}+i\delta},
\end{eqnarray}
where $f^B_{\bf{p}}$ -- Bose distribution function, $E_{\bf{p}}=p^2/2M$ -- exciton kinetic energy and $M$ is exciton mass. To calculate polarization operator $\Pi_{k\omega}$ we consider  long-wavelength limit $k<<Mv_T$, where $v_T=\sqrt{2T/M}$ is the exciton gas thermal velocity. Expanding all expressions in (\ref{eq.10}) for a small $\textbf{k}$, we get
\begin{eqnarray} \label{eq.11}
Re\,\,\Pi_{k\omega}=\frac{M}{2\pi}\int_0^\infty dx\left[1-\frac{|\eta|\theta(\eta^2-x)}{\sqrt{\eta^2-x}}\right]f'_x,\\\nonumber
Im\,\,\Pi_{k\omega}=\frac{M}{2\pi}\int_{\eta^2}^\infty dx\frac{\eta}{\sqrt{x-\eta^2}}f'_x,
\end{eqnarray}
where $f=[\exp(x-\mu/T)-1]^{-1}$, $\eta=\omega/v_Tk$, and prime means the derivative with respect to $x$. Moreover, we also can simplify $g_k$ in (\ref{eq.10}) since $kd<<1$  for typical SAW wavelength, and one has $g_{k=0}\approx 4\pi e^2d/\epsilon$.

The general integration in (\ref{eq.11}) is not possible and we will consider two limiting cases: $\eta=\omega/v_Tk=c_t\xi_0/v_T<<1$ and $\eta>>1$. These inequalities compare SAW velocity $c_t\xi_0$ with exciton gas thermal velocity $v_T$. The simple calculations yield the following results. If $c_t\xi_0/v_T<<1$
\begin{eqnarray} \label{eq.12}
\frac{\Delta c}{c}=-k\frac{\lambda^2M}{\pi\rho c_l^2}\frac{\xi_0\sqrt{1-\xi_0^2}}{f'(\xi_0)}\frac{e^{\frac{2\pi N_0}{MT}}-1}{1+\frac{2d}{a}\left(e^{\frac{2\pi N_0}{MT}}-1\right)},\\\nonumber
\Gamma=-k^2\frac{2\lambda^2M}{\pi\rho c_l^2}\frac{\xi_0\sqrt{1-\xi_0^2}}{f'(\xi_0)}\frac{B(T)c_t\xi_0/v_T}{\left[1+\frac{2d}{a}\left(e^{\frac{2\pi N_0}{MT}}-1\right)\right]^2}.
\end{eqnarray}
Here $B(T)=\int_0^\infty f'_x dx/\sqrt{x}<0$, $a=\epsilon/Me^2$ and $N_0$ is equilibrium exciton density. In the opposite case of $c_t\xi_0/v_T>>1$ we found
\begin{eqnarray} \label{eq.13}
\frac{\Delta c}{c}=k\frac{2\lambda^2}{\rho c_l^2}\frac{\xi_0\sqrt{1-\xi_0^2}}{f'(\xi_0)}\frac{N_0v_T^2}{2T(c_t\xi_0)^2},\\\nonumber
\Gamma=k^2\frac{2\lambda^2M}{\rho c_l^2\sqrt{\pi}}\frac{\xi_0\sqrt{1-\xi_0^2}}{f'(\xi_0)}\frac{c_t\xi_0}{v_T}\left(1-e^{-\frac{2\pi N_0}{MT}}\right)e^{-\left(c_t\xi_0/v_T\right)^2}.
\end{eqnarray}
We discuss these results below in the last section of the paper.

\section{$T<T_c$}
In this section we will consider the response function $S_{k\omega}$ in the presence of Bose condensate. It is known that the elementary excitations of the Bose-condensed system are Bogoliubov quasi-particles. An explicit form of dispersion law of the Bogoliubov excitations depends on the model used to describe the interacting exciton system. In the case of small exciton density $N_0a_B^2<<1$, where $a_B$ is a Bohr exciton radius, an appropriate theoretical model is the Bogoluibov model of weakly-interacting Bose gas. In the framework of this model, the dispersion law of elementary excitations has the form of $\varepsilon_k=\sqrt{\frac{k^2}{2M}\left(\frac{k^2}{2M}+2g_0n_c\right)}$. Here, $n_c$ is exciton density in the condensate. In long-wavelength limit $\frac{k^2}{2M}<<2g_0n_c$, elementary excitations are sound quanta $\varepsilon_k\approx sk$, where $s=\sqrt{g_0n_c/M}$ is its velocity. In a Bose-condensed state, most of the excitons are in the condensate, but there are also noncondensate particles, both due to the interaction and finite temperature (thermal-excited particles). These three fractions can contribute to response function $S_{k\omega}$. We will consider quantum limit $T<<sk$ when the quantum effects are the most important ones in the response function of the system to the external excitation. In quantum regime $T<<sk$ thermal excitations are not important, and the theory can be developed for $T=0$. This is the case we consider here. Due to the weak interaction between excitons, the density of the noncondensate particles is low enough, and one can disregard the interaction between the fluctuations of condensate and noncondensate densities. Thus, the response of the condensate and noncondensate particles can be calculated independently, $S_{k\omega}=S_{k\omega}^c+S_{k\omega}^n$, where $S_{k\omega}^c$ and $S_{k\omega}^n$ are response functions of the condensate and noncondensate particles, respectively.

The response of the condensate particles can be found using the Gross-Pitaevskii equation
\begin{eqnarray} \label{eq.14}
i\partial_t\Psi(\textbf{r},t)=\left(\frac{\bf{p}^2}{2M}-\mu+g_0|\Psi(\textbf{r},t)|^2\right)\Psi(\textbf{r},t)+W(\textbf{r},t)\Psi(\textbf{r},t).
\end{eqnarray}
SAW deformation field $W(\textbf{r},t)$ is treated here as a perturbation. Thus, the wave function of condensate particles $\Psi(\textbf{r},t)$ is split in the stationary uniform value and perturbed contribution $\Psi(\textbf{r},t)=\sqrt{n_c}+\psi(\textbf{r},t)$. The response function of the condensate excitons is defied as $\delta n_c(k,\omega)=S_{k\omega}^cW_{k\omega}$. Here, $\delta n_c(k,\omega)=\sqrt{n_c}(\psi^{\ast}(\textbf{r},t)+\psi^(\textbf{r},t))$ is a perturbation of the condensate particle density. Linearizing (\ref{eq.14}), we found
\begin{eqnarray} \label{eq.15}
S_{k\omega}^c=\frac{n_ck^2/M}{(\omega+i\delta)^2-\varepsilon_k^2}.
\end{eqnarray}
The calculation of the noncondensate particles response function is more cumbersome \cite{KovChapl1}, \cite{KovChapl2} and we present here the result
\begin{eqnarray}\label{eq.16}
S^{n}_{k\omega}=-\frac{g^2n_c^2}{4s^2}
\left[\frac{\theta(s^2k^2-\omega^2)}{\sqrt{s^2k^2-\omega^2}}+i\frac{\theta(\omega^2-s^2k^2)}{\sqrt{\omega^2-s^2k^2}}\right],
\end{eqnarray}

Substituting (\ref{eq.15}) and (\ref{eq.16}) in (\ref{eq.8}), we get the sound velocity renormalization and attenuation coefficient of SAW in the presence of exciton BEC
\begin{eqnarray}\label{eq.17}
\frac{\Delta c}{c}=\frac{2\lambda^2\xi_0\sqrt{1-\xi_0^2}}{f'(\xi_0)\rho c_l^2}\left[\frac{n_ck/M}{c_t^2\xi_0^2-s^2}-\frac{g^2n_c^2\theta(s^2-c_t^2\xi_0^2)}{4s^2\sqrt{s^2-c_t^2\xi_0^2}}\right],\\\nonumber
\Gamma=k\frac{4\lambda^2\xi_0\sqrt{1-\xi_0^2}}{f'(\xi_0)\rho c_l^2}\left[\frac{\pi n_ck}{2Ms}\delta(c_t\xi_0-s)+\frac{g^2n_c^2\theta(c_t^2\xi_0^2-s^2)}{4s^2\sqrt{c_t^2\xi_0^2-s^2}}\right].
\end{eqnarray}
Here, the first and second terms are the condensate and noncondensate contributions, respectively. $\theta(x)$ is a Heaviside step function.

\section{SAW-exciton interaction via piezoelectric coupling} To study the piezoelectric coupling, we must take into account the anisotropy of the substrate crystal lattice. Such an approach results in very cumbersome calculations and equations. To simplify the theoretical analysis, we will follow paper \cite{ChaplikKrasheninnikov} and will consider the mechanical motion of the substrate as the motion of isotropic medium and will include anisotropy into the piezoelectric terms of the equations of motion. Let us assume that the substrate is made of the cubic crystal and the SAW travels along piezo-active direction $[110]$ ($x$ axis in Fig.1) on surface $[001]$ ($z=0$ plane in Fig.1). In this geometry, the motion of the medium and electric field satisfy the following equations
\begin{eqnarray} \label{eq.1.1}
(\omega^2-c_l^2k^2)u_x+c_t^2u''_x+(c_l^2-c_t^2)iku'_z-2i\beta k\phi'/\rho=0,\\\nonumber
(\omega^2-c_t^2k^2)u_z+c_l^2u''_z+(c_l^2-c_t^2)iku'_x+k^2\beta\phi/\rho=0,\\\nonumber
\epsilon(z)(\phi''-k^2\phi)+8\pi\beta(iku'_x-u_zk^2/2)=0,
\end{eqnarray}
where prime means the derivative with respect to $z$. Equations of motion (\ref{eq.1.1}) must be supplemented with the boundary conditions for displacement vector components
\begin{eqnarray} \label{eq.1.2}
\rho c_t^2(u'_x+iku_z)-i\beta k\phi=0\\\nonumber
(c_l^2-2c_t^2)iku_x+c_l^2u'_z=0.
\end{eqnarray}
Moreover, the Poisson equation in the system (\ref{eq.1.1}) needs boundary conditions for electric induction vector $\textbf{D}$ and electric potential $\phi$. From the electrostatic point of view, the exciton layer can be viewed as an electric double layer. To apply this model, conditions $kd<<1$ and $k/\kappa_{t,l}<<1$ must be satisfied. The boundary conditions for the double electric layer at point $z=0$ have the form
\begin{eqnarray} \label{eq.1.3}
\phi'(+0)-\phi'(-0)=4\pi \beta iku_x(-0),\\\nonumber
\phi(+0)-\phi(-0)=4\pi pn_{k\omega}
\end{eqnarray}
Here, $p=ed$ is the absolute exciton dipole moment value and $n_{k\omega}=S_{k\omega}[-pE_z(-0)]$ is exciton density fluctuation caused by piezoelectric field $E_z(-0)=\frac{4\pi\beta}{\epsilon}iku_x(-0)$.

The general solution of eqs.(\ref{eq.1.1}) is not a simple problem. To solve it we will use the fact that the piezoeffect is a small perturbation (the mathematical criterion will be given below). Thus, the $\beta$-dependent terms in eqs.(\ref{eq.1.1}) will be considered as a perturbation. The solutions of eqs.(\ref{eq.1.1}) can be presented in the form
\begin{eqnarray} \label{eq.1.4}
u_x(z)=u^0_x(z)+\delta u_x(z),\\\nonumber
u_z(z)=u^0_z(z)+\delta u_z(z),\\\nonumber
\phi(z)=\phi^0(z)+\delta\phi(z),
\end{eqnarray}
where unperturbed functions $u_x^0(z),\,\,u^0_z(z)$ are given by eq.(\ref{eq.2}), and the unperturbed potential is
\begin{eqnarray} \label{1.5}
\phi^0(z)=Ce^{-kz}, & z>0 \\\nonumber
\phi^0(z)=De^{kz}, & z<0.
\end{eqnarray}
The corrections to unperturbed solutions can be found from eqs.(\ref{eq.1.1}) in the first order in $\beta$
\begin{eqnarray} \label{1.6}
\delta\phi(z)=xAe^{\kappa_tz}+yBe^{\kappa_lz}\\\nonumber
\delta u_x(z)=\eta De^{kz}\\\nonumber
\delta u_y(z)=\xi De^{kz}
\end{eqnarray}
where the coefficients are given by
\begin{eqnarray} \label{1.7}
\left(
  \begin{array}{c}
    \eta \\
    \xi \\
  \end{array}
\right)=\frac{\beta k^2}{\rho\omega^4}
\left(
  \begin{array}{c}
    2i\omega^2+3i(c_l^2-c_t^2)k^2 \\
    -\omega^2+3(c_l^2-c_t^2)k^2 \\
  \end{array}
\right),\\\nonumber
\left(
  \begin{array}{c}
    x \\
    y \\
  \end{array}
\right)=\frac{4\pi\beta}{\varepsilon}ik\left(
          \begin{array}{c}
            \frac{2\kappa_t^2+k^2}{k^2-\kappa_t^2} \\
            \frac{3k\kappa_l}{k^2-\kappa_l^2} \\
          \end{array}
        \right)
\end{eqnarray}

Substituting eq.(\ref{eq.1.4}) in the boundary conditions eqs.(\ref{eq.1.2}) and (\ref{eq.1.3}) we come to the dispersion equation
\begin{eqnarray} \label{eq.1.8}
f_{k\omega}=-\frac{4\pi p^2}{\varepsilon}\gamma kS_{k\omega}L_{k\omega},
\end{eqnarray}
where
\begin{eqnarray} \label{eq.1.9}
L_{k\omega}=\kappa_t(\kappa_t^2-k^2)\left[1-\frac{c_t^2k^2}{\omega^2}\left(1+\frac{6k^2(c_t^2-c_l^2)}{\omega^2}\right)\right]+\\\nonumber
+\frac{c_l^2k^2}{\omega^2}k\left[\frac{\omega^2}{c_l^2}+(\kappa_t-\kappa_l)^2\right]\left[3-4\frac{c_t^2}{c_l^2}\left(1+\frac{3k^2(c_l^2-c_t^2)}{2\omega^2}\right)\right].
\end{eqnarray}
On the right hand side of eq.(\ref{eq.1.8}), we kept only the excitonic contribution and neglected the terms due to the piezoeffect in the absence of excitonic gas.

The calculation of velocity renormalization and attenuation coefficient is similar to the procedure described above. We present here the results. For $c_t\xi_0<<v_T$
\begin{eqnarray} \label{eq.1.10}
\frac{\Delta c}{c}=\gamma\frac{2Mp^2}{\varepsilon}\frac{L(\xi_0)}{\xi_0f'(\xi_0)}\frac{e^{2\pi N_0/MT}-1}{1+\frac{2d}{a}\left(e^{2\pi N_0/MT}-1\right)}\\\nonumber
\Gamma=\gamma k\frac{4Mp^2}{\varepsilon}\frac{L(\xi_0)}{\xi_0f'(\xi_0)}\frac{B(T)c_t\xi_0/v_T}{\left[1+\frac{2d}{a}\left(e^{2\pi N_0/MT}-1\right)\right]^2},
\end{eqnarray}
and for the opposite case $c_t\xi_0>>v_T$
\begin{eqnarray} \label{eq.1.11}
\frac{\Delta c}{c}=-\gamma\frac{4\pi p^2}{\varepsilon}\frac{L(\xi_0)}{\xi_0f'(\xi_0)}\frac{N_0v_T^2}{2T(c_t\xi_0)^2}\\\nonumber
\Gamma=-\gamma k\frac{4Mp^2}{\varepsilon\sqrt{\pi}}\frac{L(\xi_0)}{\xi_0f'(\xi_0)}\frac{c_t\xi_0}{v_T}\left(1-e^{-\frac{2\pi N_0}{MT}}\right)e^{-\left(c_t\xi_0/v_T\right)^2},
\end{eqnarray}
where $L(\xi_0)<0$.

To find the SAW attenuation and velocity renormalization in the presence of the excitonic BEC, we use response functions (\ref{eq.15}) and (\ref{eq.16}). The simple calculations yield
\begin{eqnarray}\label{eq.1.12}
\frac{\Delta c}{c}=-\gamma\frac{4\pi p^2L(\xi_0)}{\varepsilon\xi_0f'(\xi_0)}\left[\frac{n_c/M}{c_t^2\xi_0^2-s^2}-\frac{g^2n_c^2\theta(s^2-c_t^2\xi_0^2)}{4ks^2\sqrt{s^2-c_t^2\xi_0^2}}\right],\\\nonumber
\Gamma=-\gamma\frac{8\pi p^2L(\xi_0)}{\varepsilon\xi_0f'(\xi_0)}\left[\frac{\pi n_ck}{2Ms}\delta(c_t\xi_0-s)+\frac{g^2n_c^2\theta(c_t^2\xi_0^2-s^2)}{4s^2\sqrt{c_t^2\xi_0^2-s^2}}\right].
\end{eqnarray}

\section{Conclusion}

We considered the SAW-exciton gas interaction at both above and below exciton gas condensation temperature $T_c$. It is shown that, above $T_c$ the SAW absorbtion coefficient is a monotonic function of exciton density for both deformation and piezoelectric interaction mechanisms, eqs.(\ref{eq.12}), (\ref{eq.13}) and (\ref{eq.1.10}), (\ref{eq.1.11}). In the presence of the exciton condensate at zero temperature, the absorption coefficient has a step-like dependence on the exciton density both for deformation and piezoelectric interaction mechanisms also. Indeed, according to eqs.(\ref{eq.17}) and (\ref{eq.1.12}), if $c_t\xi_0\neq s$, the noncondensate particle contribution to the SAW attenuation is non-zero at $c_t\xi_0>s$. The Bogoliubov quasi-particle velocity $s$ depends on the exciton concentration of condensate particles $n_c$ via relation $s=\sqrt{gn_c/M}$. Thus, inequality $c_t\xi_0>s$  is equivalent to  $n_c<n_c^0$, where the critical exciton density is given by $n_c^0=M(c_t\xi_0)^2/g$, and attenuation coefficient $\Gamma$ has a step-like dependence on exciton density
\begin{eqnarray} \label{eq.2.1}
\Gamma\propto\theta(n_c^0-n_c)
\end{eqnarray}
We can finally conclude that SAW will travel through the system \emph{without dissipation} if the exciton density of the condensate particle is less than some critical value $n_c^0$.

\section{Acknowledgments}
The authors thank Prof. A. Chaplik for useful discussions and acknowledge the financial support from the Russian Foundation for Basic Research (project 14-02-00135).

\begin{figure} [pH]
\centerline{\input epsf \epsfysize=4.0cm \epsfbox{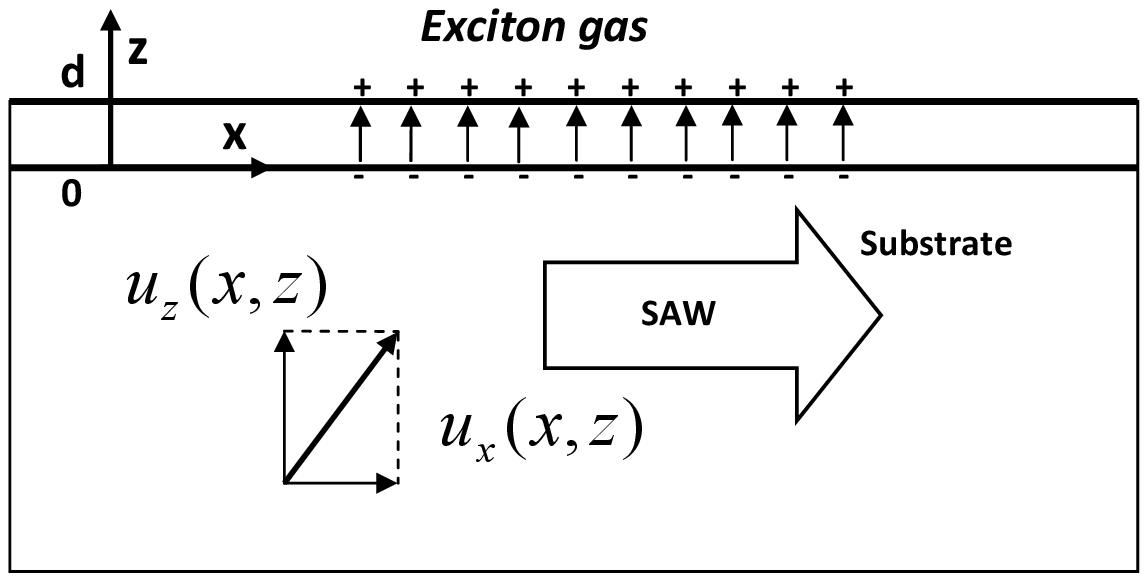}}
\caption{Sketch of the system under study.}
\label{Figure1}
\end{figure}


\begin{thebibliography}{0}

\bibitem{Moskalenko}
S.A. Moskalenko, Fiz. Tverd. Tela, \textbf{4}, {276} (1962)

\bibitem{Blatt}
J.M. Blatt, K.W. Boer and W. Brandt, {Phys. Rev.} {\textbf{126}}, {1691} (1962)

\bibitem{Keldysh_Kopaev}
L.V. Keldysh and Yu. V. Kopaev, Sov. Phys. Sol. St. {\textbf{6}}, {2219} ({1965})

\bibitem{Keldysh_Kozlov}
L.V. Keldysh and A.N. Kozlov, JETP {\textbf{27}},{521} ({1968}).

\bibitem{Keldysh_Guseinov}
R.R. Guseinov and L.V. Keldysh, JETP {\textbf{36}}, {1193} ({1973}).

\bibitem{Snoke} D. Snoke, G.M. Kauvoulakis, ArXiv:1212.4705v1

\bibitem{Snoke2}D.W. Snoke, J.P. Wolfe, A. Mysyrowicz, {Phys. Rev.}{\textbf{B41}}, {11171} ({1990}).

\bibitem{Butov}
  L.V. Butov, C.W. Lai, A.L. Ivanov, A.C. Gossard, D.S. Chemla, {Nature} {\textbf{417}}, {47} ({2002}).

\bibitem{Kasprzak}
  {J. Kasprzak, M. Richard, S. Kundermann, A. Baas, P. Jeambrun, J.M.J. Keeling, F.M. Marchetti, M.H. Szamanska, R. Andre, J.L. Staehli, V. Savona, P.W. Littlewood, B. Deveaud, L.S. Dang}
  {Nature} {\textbf{443}}, {409} ({2006}).

\bibitem{ChaplikKrasheninnikov}
 {A.V. Chaplik, M.V. Krasheninnikov}
 {Surface Science} {\textbf{98}}, {533} ({1980})

\bibitem{Landau_Lifshits}
  {Landau L.D. \and Lifshitz E.M.}
  {Theory of Elasticity}
  {7}
  {Pergamon Press, Oxford},
  {109} ({1970}).

\bibitem{KovChapl1}
 {V.M. Kovalev, A.V. Chaplik}
 {JETP Letters} {\textbf{96}}, {775} ({2013})

\bibitem{KovChapl2}
 {E.G. Batyev, V.M. Kovalev, A.V. Chaplik}
 {JETP Letters} {\textbf{99}}, {540} ({2014})

\end{thebibliography}
\end{document}